# Can the quantum spin Hall state of silicene be preserved on substrate？

Ke Yang[1], Wei-Qing Huang[1*], Wangyu Hu[2], Gui-Fang Huang[1#], Shuangchun Wen[1]

[1] *Department of Applied Physics, School of Physics and Electronics, Hunan University, Changsha 410082, China*

[2] *School of Materials Science and Engineering, Hunan University, Changsha 410082, China*

**Abstract**: The substrate-induced topological phase transition of silience is a formidable obstacle for developing silicene-based materials and devices for compatibility with current electronics by using its topologically protected dissipationless edge states. First-principles calculations indicate that the substrate will result in a phase transition from topological nontrivial phase to trivial phase of silicene, although its Dirac cone is still obvious. The substrate effect——equivalent to an electric field——annihilates its spin-orbit coupling effect, the reason why its quantum spin Hall effect (QSHE) of silicene has not been experimentally observed. Unfortunately, external electric field seems impossible to recover the QSHE due to the screen effect of substrate. We here first propose a viable strategy——constructing inverse symmetrical sandwich structure (protective layer/silicene/substrate)——to preserve quantum spin Hall (QSH) state of silicene, which is demonstrated by using two representatives ($CeO_2(111)$/silicene/$CeO_2(111)$  and $CaF_2(111)$/silicene/$CaF_2(111)$) through the calculated edge states and $Z_2$ invariant. This work takes a critical step towards fundamental physics and realistic applications of silicene-based nanoelectronic devices.

---

[*]. Corresponding author. *E-mail address:* wqhuang@hnu.edu.cn
[#]. Corresponding author. *E-mail address:* gfhuang@hnu.edu.cn

Two-dimensional (2D) materials (such as graphene and silicene) with a honeycomb geometry have attracted increasing attention owing to their unique electronic properties [1-4]. Ever since the prediction of quantum spin Hall (QSH) state in freestanding silicon monolayer (silicene), researchers have been infatuated with the idea of its applications in spintronic device using its topologically protected dissipationless edge states, because of its uniquely suitable for integration in Si-based electronics. Compared to graphene, silicene possesses a large intrinsic spin-orbit coupling (SOC), opening a band gap (~1.6 meV[5]) at Dirac points and topological nontrivial electronic structure, which can host the quantum spin Hall effect (QSHE) up to 18 K[5]. However, in sharp contrast to graphene, which can be mechanically exfoliated from graphite, silicon atoms in equilibrium low-buckled silicene adopt $sp^2/sp^3$-mixed hybridization states [6], making it difficult to obtain free-standing monolayer silicene.

To realize the silicene spintronic devices based on QSH state, there are, until now, three essential ongoing challenges. First, an appropriate substrate to stabilize monolayer silicene is required. Second, as described below, interaction between silicene and the substrate must be weak to guarantee the existence of the Dirac cone and QSH state in silicene over an appreciable temperature range. Finally, silicene must be isolated or encapsulated to hinder its oxidation because of its air stability issue [7].

Appropriate substrates, such as Ag[8-10], Ir[11], Ru[12], ZrB2[6] and ZrC[13], for epitaxial growth of silicene have been hotly pursued both theoretically and experimentally. However, the interaction between silicene and these substrates is so strong that it markedly depress QSH state[14]. For instance, the orbital hybridization between Ag and Si atoms results in a surface metallic band and depresses the Dirac fermion characteristics [10] in an epitaxial silicene on an Ag(111) surface[15,16]. In principle, eliminating or minimizing substrate effects could preserve Dirac cone of silicene [17-20]. Until now, it is proposed that some substrates (such as $MgX_2$ (X=Cl, Br and I) [21], GaS[22], $CaF_2$ [18], and BN [23, 24]) can approximately preserve Dirac cone of silicene. Particularly, in the ingenious experiment, ARPES measurements have demonstrated Dirac cone of quasi-freestanding silicene by oxygen intercalation to weaken interaction with its Ag(111) substrate [25]. However, compared to freestanding silicene, the band gap in Dirac cone of silicene on the substrates are significantly large [18, 21-23, 25, 26], due to the equivalent electric field induced by substrate [23, 27]. Unfortunately, the substrate-induced electric field makes silicene occurring topological phase transition from QSHE to band insulator [28]. This seems to be the reason why the QSH state in silicene on these substrates has, until now, few theoretical investigation, let alone experimental observations. Therefore, the fundamental question naturally arises: Can the QSH state of silicene be preserved on substrate at realistic temperature?

In this Letter, we elucidate the correlation between topological properties, stability and binding energy of silicene with its substrate with particular attention to how to preserve its QSH state. Based on first principle calculations, we find that silicene has become a band insulator regardless of how weak the interaction as it is steadily grown on substrate. The quantitative relation between binding energy and band gap reveals that the QSH state is almost impossible to be preserved in silicene/substrates system, the most likely reason why no QSHE of silicene have been demonstrated experimentally. We propose a viable strategy, by forming inverse symmetrical sandwich structure (protective layer/silicene/substrate) to preserve QSH state of silicene. This is of particular technological interest because it is the required prerequisite to generate Hall current of spin charges on the edges of the silicene in integrated spintronic devices.

Density functional theory (DFT) was preformed to achieve optimized geometrical and electronic structures with a projector augmented wave (PAW)[29] basis as implemented in Vienna Ab Initio Simulation Package code[30, 31]. The Perdew-Burke-Ernzerhof (PBE) Generalized Gradient Approximation (GGA) exchange-correlation functional method was adopted. All calculations were performed using the DFT/GGA+U method (U= 9.0 and 4.5 eV for Ce 4f and O 2p, respectively) to obtain band gap of $CeO_2$. Moreover, the spin orbit coupling (SOC) was taken into account and the van der Waals interaction (DFT-D3 method with Becke-Jonson damping) was incorporated [32, 33]. The kinetic energy cutoff was 500 eV. Brillouin zone integration was performed on grids of 15 × 15 × 1 Monkhorst–Pack k-points. Total energy and all forces on atoms converged to less than $10^{-8}$ eV and 0.005 eV/Å. The vacuum space of 20 Å along the z direction is used to decouple possible periodic interactions.

The $CeO_2$ (111) surface is first chosen as a candidate substrate of epitaxial silicene. The choice of $CeO_2$ is owing to their little lattice mismatch (the lattice constant are 3.85 and 3.83 Å for freestanding silicene and $CeO_2$(111) surface, respectively; thus the (1×1) $CeO_2$(111) surface is perfectly lattice-matched to the Bravais lattice of (1×1) silicene), avoiding the stress of local unequivalent Si atoms. Moreover, it is a high dielectric oxide protective layer in silicon-based device, in which Si(111) surface has widely used to epitaxial grow $CeO_2$ film [34-37]. By preforming extensive search for the silicene on $CeO_2$(111) surface, the optimized four different stacking patterns with high symmetry fall into two general categories: covalent and noncovalent. As the down half-layer Si atoms are positioned on top of the topmost O atoms, the covalent bond between Si and O atoms is formed. The chemically bonded silicene/$CeO_2$(111) system can be divided CA configuration (the upper layer Si atoms are located on top of the second O atoms) in Fig. 1(a1) and CB configuration (the upper layer Si atoms are located on Ce atoms) in Fig. S1 (a1). For the noncovalent silicene/$CeO_2$(111) system, the upper layer Si atoms positioned on top of Ce atoms is denoted as NA configuration (Fig. 1(b1)), while the lower layer Si atoms positioned on top of Ce atoms is named as NB configuration (Fig. S1 (b1)). Ab-initio molecular dynamics simulations and phonon spectrum calculations of four configurations reveal that these structures are thermodynamics stable, and the details will be systematically discussed elsewhere.

To explore the topological properties of silicene on $CeO_2$(111), the band structures have been calculated by DFT. Due to the strong chemical bond between Si and O atoms, the orbital of Si atom crosses the Fermi level and Dirac cone disappears in CA and CB configurations (Figs. 1(a2) and S1(a2)), thus the QSH state would not be preserved. As expected, the approximate linear Dirac cone of silicene can be still preserved in NA (Fig. 1(b2)) and NB (Fig. S1(b2)) configurations. Moreover, Dirac cone states in silicene near Fermi level are far from (about 1.5 eV in energy) the valence band maximum and conduction band minimum of $CeO_2$, demonstrating that $CeO_2$ is not only an appropriate substrate for growth of silicene [26], but also would not destroy its Dirac cone. Meanwhile, a large band gap of silicene is opened at Dirac cone (245.2 and 286.1 meV for NA and NB configurations, respectively, Table I), just as silicene is grown on other substrates [19, 26, 38].

To reveal the physical origin of band gap increasing, we have constructed the low-energy effective model of silicene on substrates presenting of SOC near the K points by tight-binding methods [28, 39]:

$$H_k = \hbar v_F(\sigma_x k_x + \sigma_y k_y) + \gamma_{so}\sigma_z s_z + \frac{\gamma_{R1}}{2}(\boldsymbol{\sigma} \times \mathbf{s}) + \gamma_m s_z + \gamma_B \mathbf{s}_z \qquad (1)$$

where $v_F$ is Fermi velocity. $\boldsymbol{\sigma}$ and $\mathbf{s}$ are the Pauli matrices, where $\boldsymbol{\sigma}$ represents the A(B)-sublattice of silicene, and $\mathbf{s}$ represents the spin. $\gamma_{so}$ and $\gamma_{R1}$ are the intrinsic and Rashba spin-orbit terms, respectively. $\gamma_m$ and $\gamma_B$ represent the "mass" and "pseudomagnetic" term. These parameters can easily derived by the calculated energy at K point and are listed in Table I. More details are given in Supplementary Materials.

The band structure near the Dirac point from the low-energy effect Hamiltonian is quite close to the results of DFT (see Figs. 2(a) and S2 (a)), indicating that the tight-binding model can describe the electronic properties of silicene on substrates well. Clearly, the band gap of silicene on $CeO_2$ is dramatically increased compared with that of free-standing silicene, which can mainly be attributed to the "mass" terms ($\gamma_m$) appearing. The large "mass" terms (119.6 and 142.1meV for NA and NB configurations, respectively; Table I) imply that the substrate effect is equivalent to a perpendicular electric field, which its magnitude depends on stacking patterns. Direct DFT calculation shows that the static electronic potential step between upper and lower layer Si atoms are 231.1 and 295.4 meV for NA and NB configurations, in well agreement with the sublattice potential step ($2\gamma_m$) by tight-binding model. Unexpectedly, the spin-orbit term is changed due to substrate effect: its magnitude relies on the buckle height of silicene and stacking patterns; in particular, the bigger buckle height (or sublattice), the stronger the spin-orbit interaction will be [27, 39]. Compared with "mass" terms, however, the change of spin-orbit term is slight. To compare the intrinsic "atomic" SOC term of monolayer silicene and the SOC induced ($\gamma$) by $CeO_2(111)$ surface, we construct ideal silicene/$CeO_2(111)$ models, which are only varied the interfacial distance and without optimizing structure of silicene (named as NA-H or NB-H configuration). The calculated $\gamma_{so}^{ind}(=\gamma_{so}^{h-NA} - \gamma_{so}^{silicene})$ are 0.67 and 0.20 $meV$ for NA and NB configurations, respectively; indicating the substrate increases the intrinsic SOC. Due to breaking of inversion symmetry, the $CeO_2(111)$ surface also induces the extrinsic Rashba SOC, which its strength (Table I) is positive correlation with the sublattice potential and intrinsic SOC, in agreement with others[27]. A "pseudomagnetic" term, induced by $CeO_2(111)$ surface, is too weak to influence the topological properties of silicene. Obviously, $\gamma_m$ is much larger than the other three parameters. Therefore, the large equivalent electric field (the "mass" term) induced by substrate is the main cause of increasing band gap of silicene.

It has been demonstrated that the system will be a topological insulator if the $\gamma_{so}$ term is dominant, whereas if $\gamma_m$ is dominant, the system is a normal insulator [3, 28]. To discuss the topological properties of silicene on $CeO_2$ (111) surface, we have calculated its edge states and $Z_2$ invariant. Figs. 2(b) shows the band structure of zigzag nanoribbon of silicene on $CeO_2$ (111) surface (NA configuration). One can see the valence band is completely full, and there are no crossing point between valence band and conduction band. The electron in valence band needs extra energy to transfer to conduction band, suggesting that silicene on $CeO_2$ (111) surface a normal band insulator. Similarly, the transition from QSH state to band insulator induced by extra electric field has been discussed in graphene[40]. Moreover, the topological trivial phase can be verified by $Z_2$ invariant. Considering the breaking of inversion symmetry in silicene/substrates, $Z_2$ invariant has been calculated by using the non-Abelian Berry connection[41], based on the evolution of the charge centers of the wannier functions, and implemented in the WannierTools code[42]. One can easily get $Z_2$=0 for NA and NB configurations (Figs. 2(c) and S2(c)). This indicates that the $CeO_2$ (111) surface induces topological phase transition of silicene from QSHE to band insulator, although the binding energy is very weak and Dirac cone of silicene is also approximately preserved.

The topological phase transition induced by substrate can also be verified by silicene put on $CaF_2$ (111) surface. The natural cleavage $CaF_2$ (111) surface has a small lattice mismatch to silicene, resulting into six different stacking patterns with high symmetry (Fig. S3). The band structures of silicene/$CaF_2$ have shown in Figs. S4 and S5, in which Dirac cone of silicene are approximately preserved. Similar with $CeO_2$, $CaF_2$ (111) surface also induces a large band gap at Dirac cone of silicene, although the interaction between $CaF_2$ (111) and silicene is very weak (Table SI). Similarly, a large band gap of silicene is also induced by monolayer BN (Fig. S12). Using tight-binding methods, the increased band gap of silicene is also attributed to the equivalent electric field induced by $CaF_2$ (111) surface. Thus, the silicene/$CaF_2$ system is also a normal insulator because $\gamma_m$ is dominant (Table SI), and $Z_2$ invariant also supports the transition from the topologically non-trivial nature to topologically trivial insulator.

These results and others [24] have confirmed that the substrates always induce the topological phase transition of silicene. In principle, the QSH state of silicene can be preserved as the substrate interaction is small enough. However, the interaction between silicene and substrate should be strong enough to stabilize silicene. To deal with this dilemma, the key issue is to expound the quantitative relationship between binding energy and band gap, and topological properties of silicene on substrate. For van der Waals heterostructures, the bigger the interfacial spacing, the weaker the interaction is. Fig. 3 displays the evolution of band gap and binding energy ($E_b = (E_{tot} - E_{silicene} - E_{substrate})/N$, where N is the number of Si atoms) with interfacial distance of silicene/$CeO_2$ (111). Evidently, the band gap of silicene is firstly decreased rapidly with the interfacial distance (*d*) increasing from equilibrium spacing, indicating from another perspective that the interaction with substrate will result in an increase of band gap of silicene. Further increasing of the interfacial distance, its band gap decreases slowly, and then gradually increases to the value (1.6 meV) of freestanding silicene, thus creating a crossover. The evolution of $E_g$ suggests the appearance of semimetal phase ($E_g$=0) [41], when the interfacial distance reaches to a critical value ($d_c$: $d_c \in$(4.60, 4.80 Å) and (5.18, 5.38 Å) for NA and NB configurations, respectively). The semimetal phase is a critical phase, implying the transition from band insulator (d<$d_c$) to QSH state (d>$d_c$). As expected, Fig. 3 demonstrates that the binding energy between silicene and $CeO_2$ (111) surface is continuously reduced as the interfacial distance increases. As a consequence, to preserve the QSH state of silicene on $CeO_2$ (111) surface, their interfacial distance should be large enough, where their binding energy is small (about 30 meV), roughly equivalent to typical thermal energy at room temperature. This demonstrates that preserving both the QSH state and the stability of silicene is mutually exclusive on substrate, due to the fact that the Dirac cone consisted of $p_z$ states is easily destroyed[43].

Since the substrate effect on the electronic structure of silicene can be equivalent to an electric field, the natural question is: whether can its QSH state be recovered by directly applying an external electric field? We calculate the band structures of silicene/$CeO_2$ (111) (NA and NB configuration) under different external electric fields, as given in Figs. S6 and S7. Disappointingly, the large band gap of silicene could hardly be tuned by external electric field, regardless of its magnitude and direction. This is due to the fact that the substrate may largely screen the extra electric field [43]. Even worse, the strong electric field will make Dirac cone of silicene into the valence or conduction band of $CeO_2$, implying the QSH state cannot be recovered by directly applying an external electric field.

To facilitate its applications in spintronic device, we here first propose a practical strategy——constructing inverse symmetrical sandwich structure (protective layer /silicene/substrate) —— to preserve the QSH state of silicene. The $CeO_2$ (111)/silicene/$CeO_2$ (111) structures (Fig. S8) with inverse symmetry are constructed to demonstrate this method. In this kind of structure, the $CeO_2$(111) surface is both substrate and protective layer. DFT calculation shows that their Dirac cone is very close to that of freestanding silicene (Figs. 4 (a) and S9 (a)), and their band gaps (0.6 and 3.0 meV for trilayer NA and NB configurations, respectively) are much smaller than those of silicene/$CeO_2$(111). For trilayer structure, the extra Rashba SOC $\gamma_{R1}$, "mass" term $\gamma_m$ and "pseudomagnetic" term $\gamma_B$ are all vanished. Band structures of zigzag nanoribbon of $CeO_2$(111)/silicene/$CeO_2$(111) show the crossing of the edge states at the Brillouin zone boundary (Figs. 4 (b) and S9 (b)). Moreover, $Z_2$ invariant obtained by evolution of charge centers of the wannier functions (Figs. 4 (c) and S9(c)) reveals that the $CeO_2$(111)/silicene/$CeO_2$ (111) is in the topological nontrivial phase.

The universality of this strategy can further be verified by $CaF_2$ (111)/silicene/$CaF_2$ (111), in which six configurations have been taken into account (Fig. S10). One can see from Fig. S11 that their Dirac cones with linear dispersions are clear, and their band gaps are much smaller than that of silicene/$CaF_2$ (111). Meanwhile, $Z_2$ invariant also reveals that they are in the topological nontrivial phase. More importantly, such symmetrical sandwich structure (protective layer/silicene/substrate) can not only preserve the QSH state, but also protect its stability of silicene. Recent developments in synthesis of 2D transition metal oxide nanosheets[44] and high level transfer techniques for 2D materials[45] make the integration of silicene into devices technologically feasible, including the fabrication of symmetrical sandwich structure (protective layer/silicene/substrate) proposed here.

In conclusion, we have demonstrate that the substrate interaction will destroy the QSH state of silicene, regardless of how weak the interaction, when the silicene is stable in silicene/substrate system at room temperature. The transition of silicene from the topological nontrivial phase to normal insulator has verified by edge state and topological invariant due to the substrate effect. Directly applying an external electric field seems most impossible to recover the QSH state of silicene because of the screen effect of the substrate. We first propose the inverse symmetrical sandwich structure, as an effective strategy, to preserve the QSH state, as well as the stability of silicene. This could be an important step toward development of silicene-based nanoelectronic devices.

**Acknowledgements**

The authors are grateful to the National Natural Science Foundation of China (Nos.51772085 and 11574079).

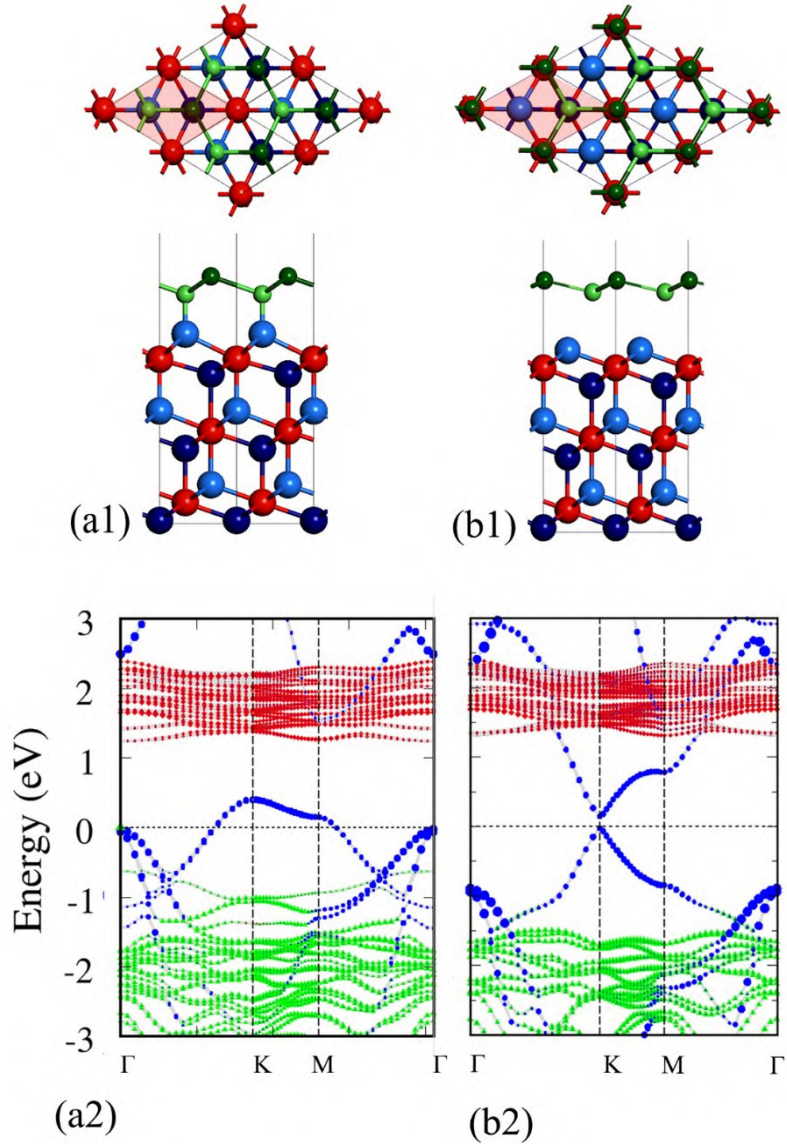

FIG. 1. Geometry (a1) for CA (the upper and lower layer Si atoms are located on top of the second and upper layer O atoms, respectively) and (b1) for NA (the upper and lower layer Si atoms are located on top of Ce and second layer O atoms, respectively) configuration. Red spheres represent Ce atoms, dark green and light green spheres represent the upper and lower layer Si atoms of silicene, and dark blue and light blue spheres represent O atoms. Band structures (a2) for CA and (b2) for NA configuration. Red, green and blue symbols represent the projected band dispersion of Ce, O and Si, respectively.

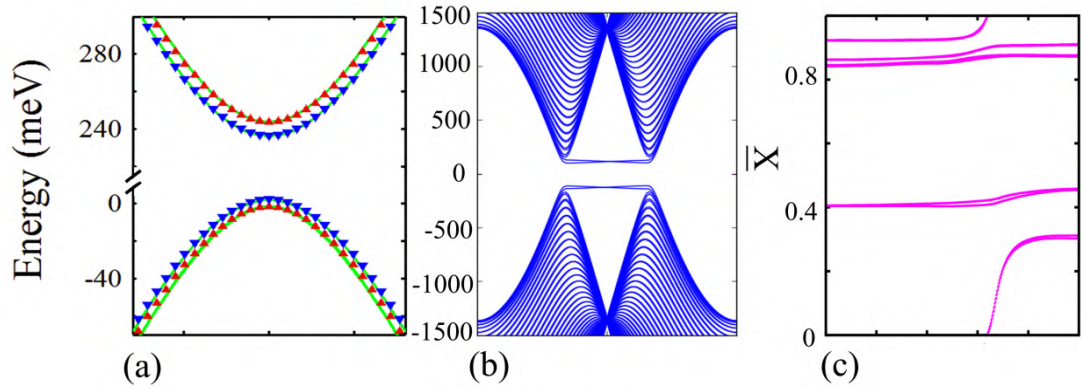

FIG. 2. (a) Band structure near K point of NA configuration. Green solid lines denote the result from tight-binding methods, triangle symbols are results of first-principles calculations. (b) Band structures of zigzag nanoribbon of silicene on $CeO_2$ (111) corresponding to NA configuration. There are no edge crossing to Fermi level. (c) The evolution of the charge centers of the Wannier functions of NA configuration, implying the $Z_2$ invariant is zero.

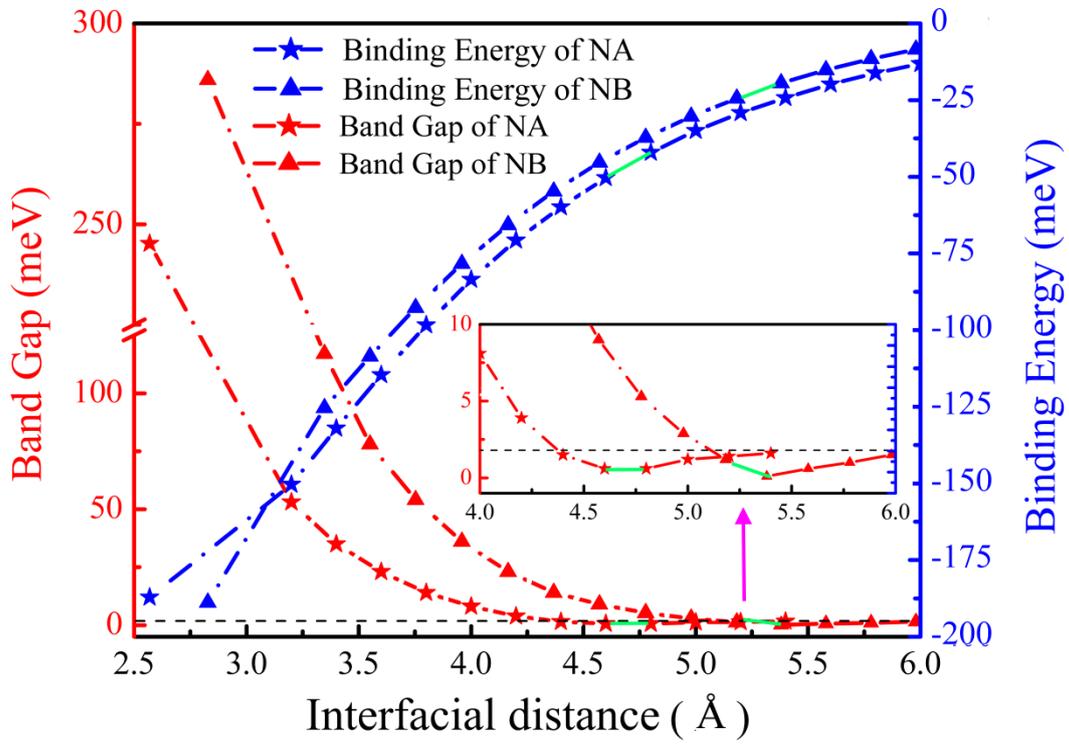

FIG.3. Band gap (red line) and binding energy (blue line) as functions of interfacial distance between silicene and $CeO_2$ (111) surface. Dot lines represent the band insulator, while the quantum spin Hall effect appears in solid line. Black dash line represents the band gap of freestanding silicene. The green solid lines imply the topological phase transitions zone, and the semimetal state (band gap is zero) appear at critical value.

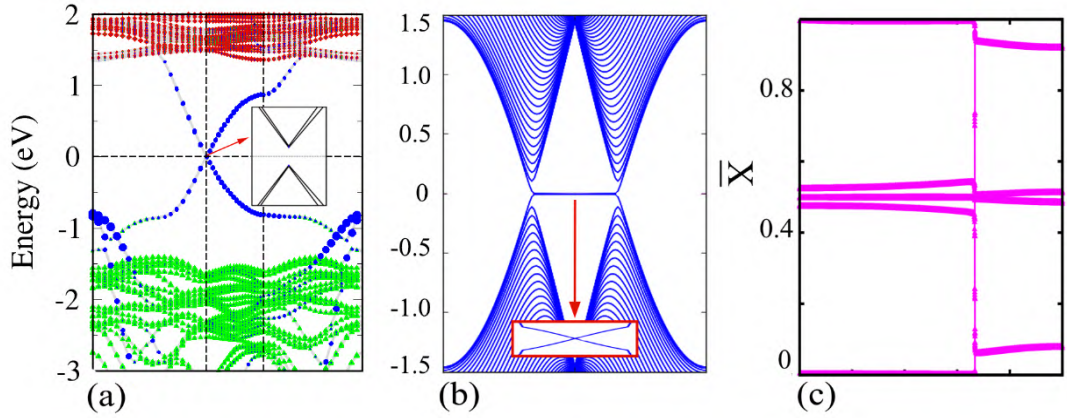

FIG.4. (a) Band structure of inverse symmetrical sandwich structure $CeO_2$ (111)/Silicene/$CeO_2$ (111) corresponding to NA configuration. Red, green and blue symbols represent the projected band dispersion of Ce, O and Si. (b) Band structure of zigzag nanoribbon of silicene on $CeO_2$ (111) corresponding to NA trilayers. (c) The evolution of the charge centers of the Wannier functions of NA trilayers, implying the $Z_2$ invariant is one.

TABLE I. Silicene on the CeO$_2$ (111) surface. The E$_b$ is the binding energy per Si atom. The E$_g$ is the gap calculated at the K point. The Hamiltonian parameters defined in Eqs. (1) are given in meV. The b is the separation between the upper and lower layer silicene. The d is the interfacial distance between silicene and substrate.

|  | E$_b$ (meV) | E$_g$ (meV) | $\gamma_m$ | $\gamma_{so}$ | $\gamma_{R1}$ | $\gamma_B$ | b (Å) | $v_F$ ($10^5$m/s) | d (Å) |
|---|---|---|---|---|---|---|---|---|---|
| Silicene | / | 1.6 | / | 0.80 | / | / | 0.47 | 5.58 | / |
| NA | 374.3 | 245.2 | 119.59 | 2.78 | 10.62 | 1.16 | 0.52 | 4.55 | 2.57 |
| NA-H | 345.2 | 183.9 | 90.43 | 1.47 | 5.72 | 0.17 | 0.47 | 5.03 | 2.59 |
| NB | 377.7 | 286.1 | 142.06 | 0.95 | 4.51 | -0.36 | 0.46 | 4.86 | 2.83 |
| NB-H | 368.7 | 287.0 | 142.48 | 1.00 | 4.70 | -0.38 | 0.47 | 4.98 | 2.83 |

# Can the quantum spin Hall state of silicene be preserved on substrate？

Ke Yang[1], Wei-Qing Huang[1*], Wangyu Hu[2], Gui-Fang Huang[1#], Shuangchun Wen

[1] *Department of Applied Physics, School of Physics and Electronics, Hunan University, Changsha 410082, China*

[2] *School of Materials Science and Engineering, Hunan University, Changsha 410082, China*

## Tight-binding methods:

To reveal the origin of band gap increasing, we have constructed the four band Hamiltonian of silicene on substrates presenting of SOC by tight-binding methods [1, 2]:

$$H = -t\sum_{<i,j>\alpha} c_{i\alpha}^\dagger c_{j\alpha} + i\gamma_{so}\sum_{\ll i,j\gg \alpha\beta} v_{ij} c_{i\alpha}^\dagger \sigma_{\alpha\beta}^z c_{j\beta} + i\frac{\gamma_{R1}(E_z)}{2}\sum_{<i,j>\alpha\beta} c_{i\alpha}^\dagger (s\times \hat{d}_{ij})_{\alpha\beta}^z c_{j\beta} - i\gamma_{R2}\sum_{\ll i,j\gg \alpha\beta} \mu_{ij} c_{i\alpha}^\dagger (s\times \hat{d}_{ij})_{\alpha\beta}^z c_{j\beta} + i\gamma_m \sum_{<i,j>\alpha\beta} \mu_i c_{i\alpha}^\dagger c_{j\alpha} + \gamma_B \sum_{i\alpha\beta} c_{i\alpha}^\dagger \sigma_{\alpha\beta}^z c_{j\beta} \quad (S1)$$

where $c_{i\alpha}^\dagger$ ($c_{j\beta}$) creates $c_{j\beta}$ an electron with spin polarization at $\alpha$ site $i$, and $<i,j>/\ll i,j\gg$ run over all the nearest or next-nearest neighbor hopping sites. The first term represents the usual nearest-neighbor hopping. The second term represents the effective SOC that contains the intrinsic "atomic" SOC term of monolayer silicene plus $\gamma_{so}^{ind}$ which can been induced by the substrate, where $s = (s_x, s_y, s_z)$ is the Pauli matrix of spin, with $v_{ij} = \pm 1$ is clockwise or anticlockwise of next-nearest-neighboring hopping with respect to the positive z axis. The third term represents the first Rashba SOC, which is induced by external electric field or substrates, and $\hat{d}_{ij} = d_{ij}/|d_{ij}|$ with the vector $d_{ij}$ connecting two sites $i$ and $j$ in the same sublattice. The forth term represents the second Rashba SOC associated with the next-nearest neighbor hopping term, where $\mu_{ij} = \pm 1$ for the A (B) site, this term is negligible for silicene, so we set $\gamma_{R2} = 0$ here. The fifth term is the staggered sublattice potential term, which describes the breaking of the sublattice symmetry by the interaction with the substrates. The sixth term represents the "pseudomagnetic" term. We expand the TB Hamiltonian surrounding the two valley K points, and obtain the low-energy effective model:

$$H_k = \hbar v_F(\sigma_x k_x + \sigma_y k_y) + \gamma_{so}\sigma_z s_z + \frac{\gamma_{R1}}{2}(\boldsymbol{\sigma}\times \boldsymbol{s}) + \gamma_m s_z + \gamma_B s_z \quad (S2)$$

---

*. Corresponding author. *E-mail address:* wqhuang@hnu.edu.cn
#. Corresponding author. *E-mail address:* gfhuang@hnu.edu.cn

where $v_F = \frac{\sqrt{3}t}{2}$, it are listed Table I. $\boldsymbol{\sigma}$ is the Pauli matrices of AB- sublattice.

From the Eq. (2), the parameters are:

$$\gamma_m = \frac{(\varepsilon_4 - \varepsilon_3) + s(\varepsilon_2 - \varepsilon_1)}{4}$$

$$\gamma_{R1} = \pm\frac{(\varepsilon_2 - \varepsilon_1)\sqrt{1-s^2}}{2}$$

$$\gamma_{so} = \frac{(\varepsilon_4 + \varepsilon_3) - (\varepsilon_2 + \varepsilon_1)}{4}$$

$$\gamma_B = \frac{(\varepsilon_4 - \varepsilon_3) - s(\varepsilon_2 - \varepsilon_1)}{4}$$

where $s$ is the expectation values for the z component of spin, and $\varepsilon$ are the eigenvalues at K point, which can been acquired by the wavefunction from DFT[2]. These parameters can easily derived by calculated the energy at K point.

# Figures:

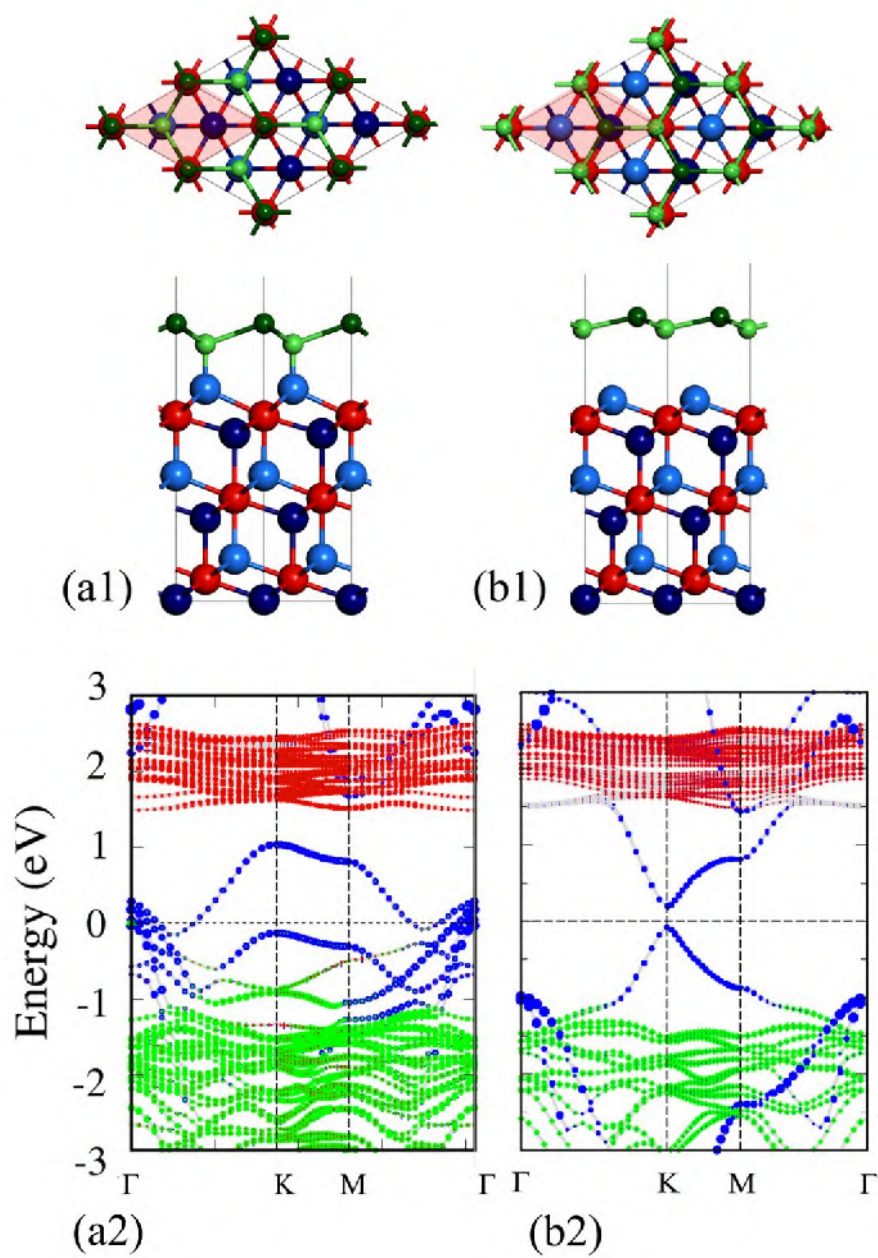

FIG. S1 Geometry (a1) for CB (the upper and lower layer Si atoms are located on top of Ce and up layer O atoms, respectively) and (b1) for NB (the upper and lower layer Si atoms are located on second layer O and Ce atoms, respectively) configuration. Red spheres represent Ce atoms, dark green and green blue spheres represent the upper and lower layer Si atoms of silicene, and dark blue and light blue spheres represent O atoms. Band structures (a2) for CB and (b2) for NB configuration. Red, green and blue symbols represent the projected band dispersion of Ce, O and Si, respectively.

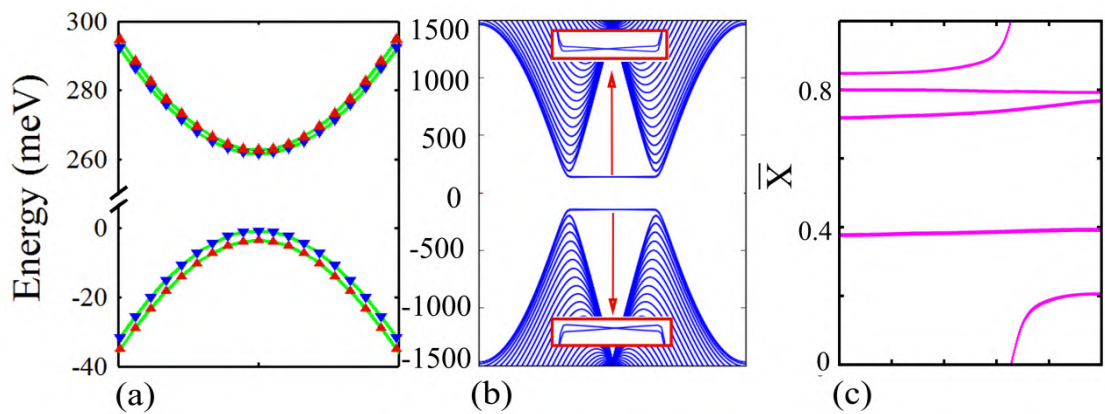

FIG. S2 (a) Band structure near K point of NB configuration. Green solid lines denote the result from tight-binding methods, triangle symbols are results of first-principles calculations. (b) Band structure of zigzag nanoribbon of silicene on $CeO_2$ (111) corresponding to NB configuration. There are no edge crossing to Fermi level. (c) The evolution of the charge centers of the Wannier functions of NB configuration, implying the $Z_2$ invariant is zero.

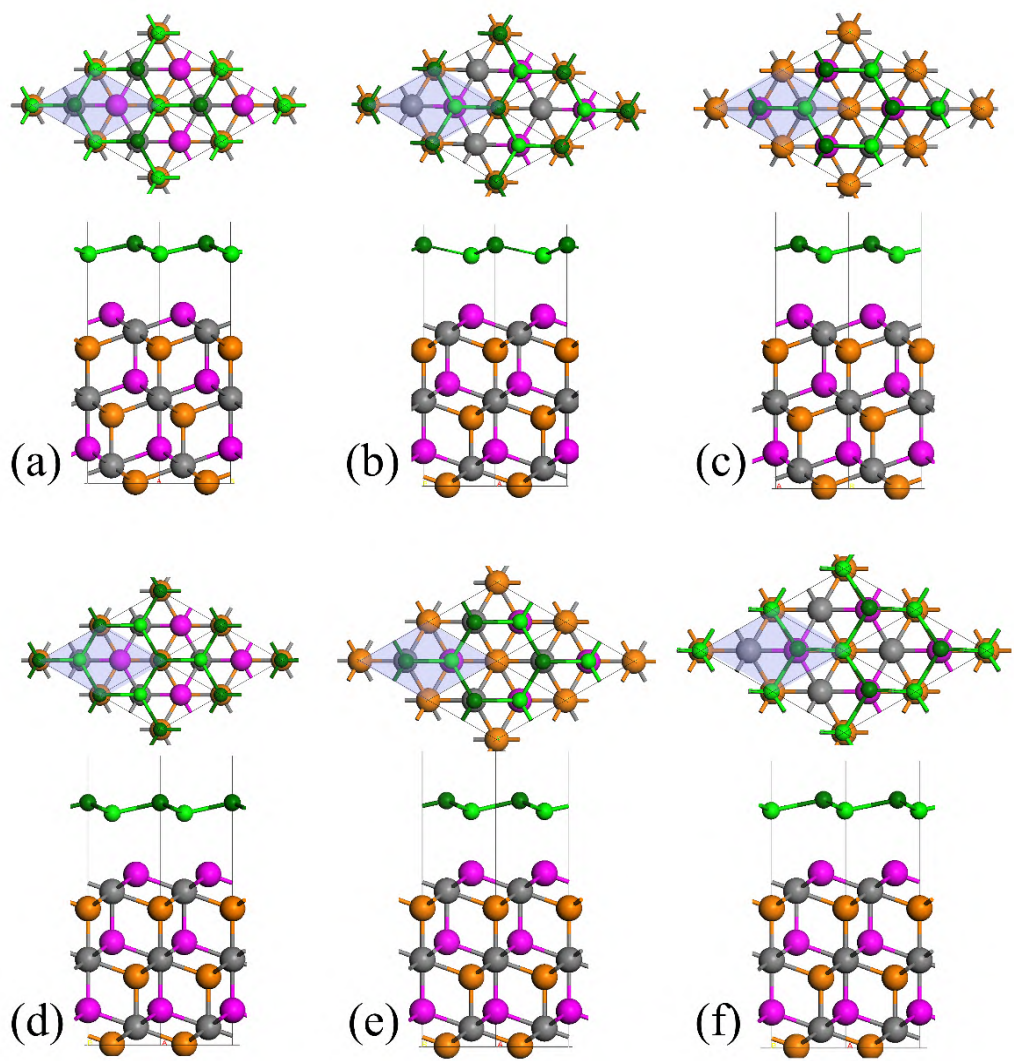

FIG. S3 Geometry for six different configurations of silience/CaF$_2$(111). Gray spheres represent Ca atoms, dark blue and light blue spheres represent up and down half-layer Si atoms of silicene, and pink and orange spheres represent F atoms.

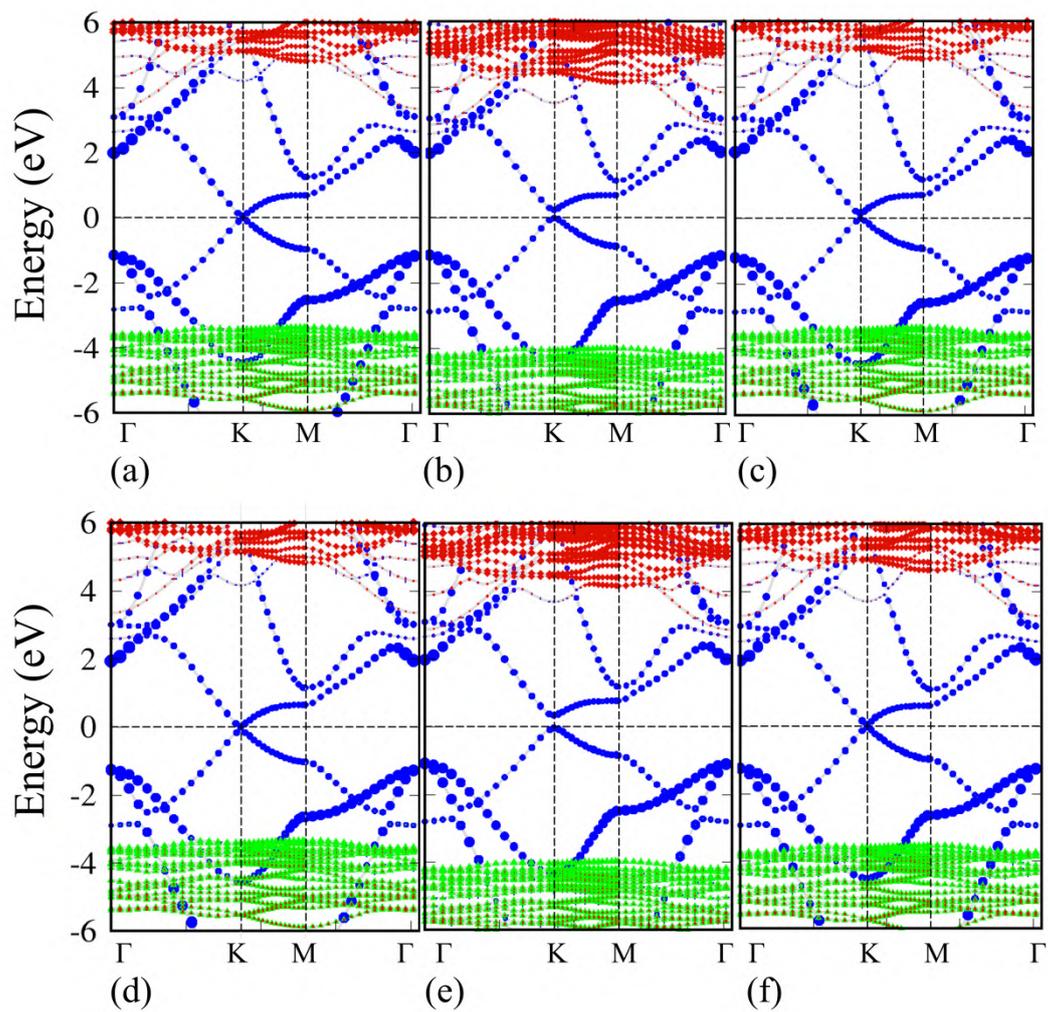

FIG. S4 (a-f) Band structures for silicene/CaF$_2$ (111) corresponding to six configurations (a-f) in Fig. S3. Red, green and blue dot represent the projected band dispersion of Ca, F and Si.

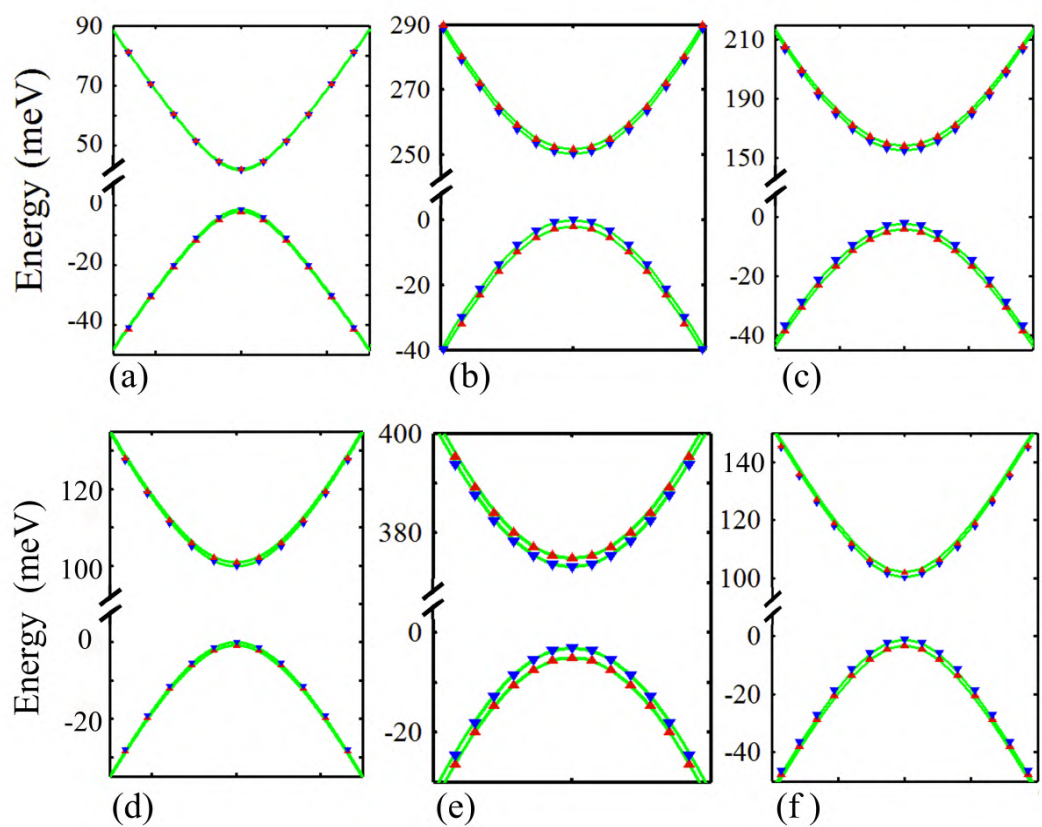

FIG. S5 (a-f) Band structures near K point of six configurations of silicene/CaF$_2$ (111) corresponding to six configurations (a-f) in Fig. S3. Green solid lines denote the result from tight-binding methods, triangle symbols are results of first-principles calculations.

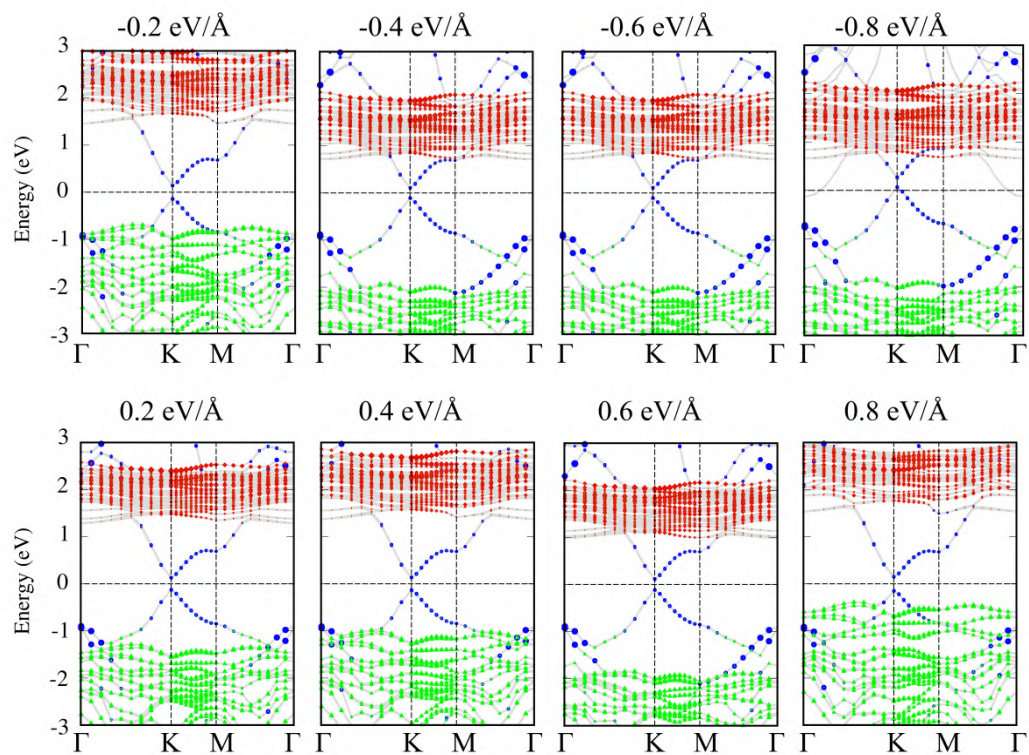

FIG. S6 The band structures of silicone/$CeO_2$(111) corresponding to NA configuration under various vertical external fields. Red, green and blue symbols represent the projected band dispersion of Ce, O and Si. Here, positive value denotes the direction of electric field from $CeO_2$ (111) to silicene, while negative value presents reverse direction of electric field.

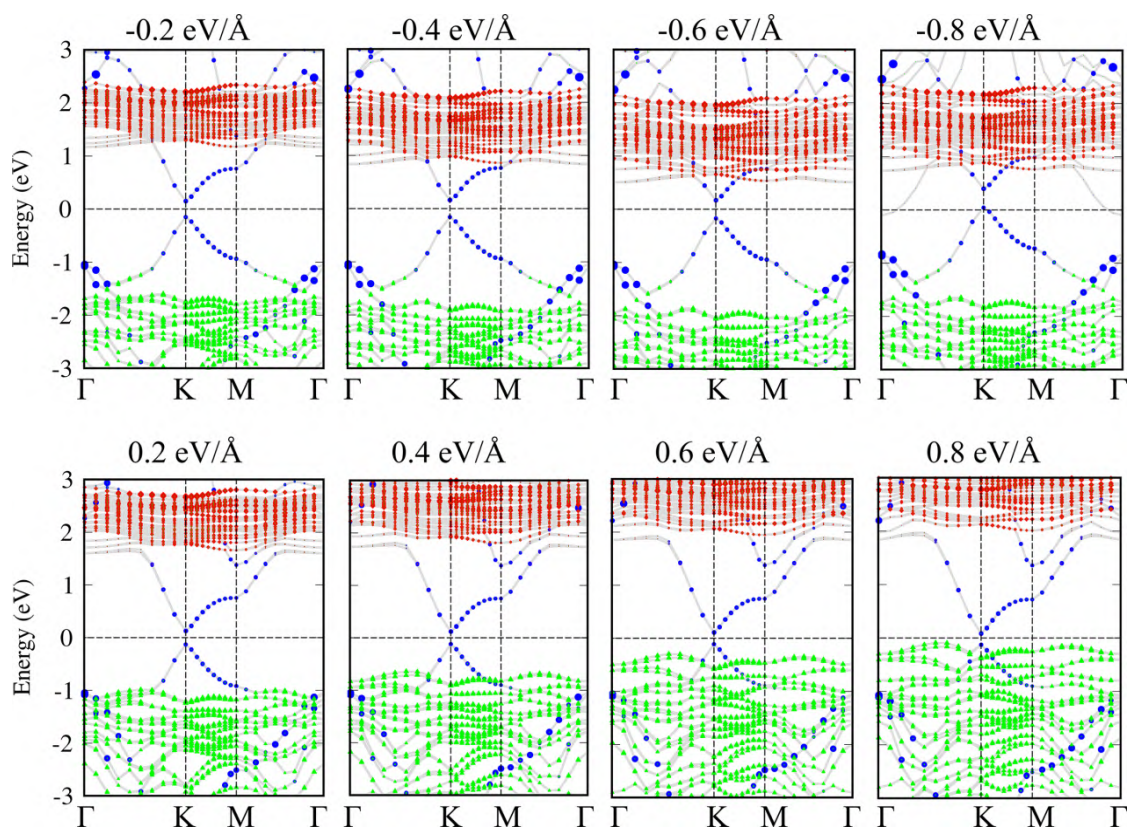

FIG. S7 The band structures of silicene/CeO$_2$(111) corresponding to NB configuration under various vertical external fields. Red, green and blue symbols represent the projected band dispersion of Ce, O and Si. Here, positive value denotes the direction of electric field from CeO$_2$ (111) to silicene, while negative value presents reverse direction of electric field.

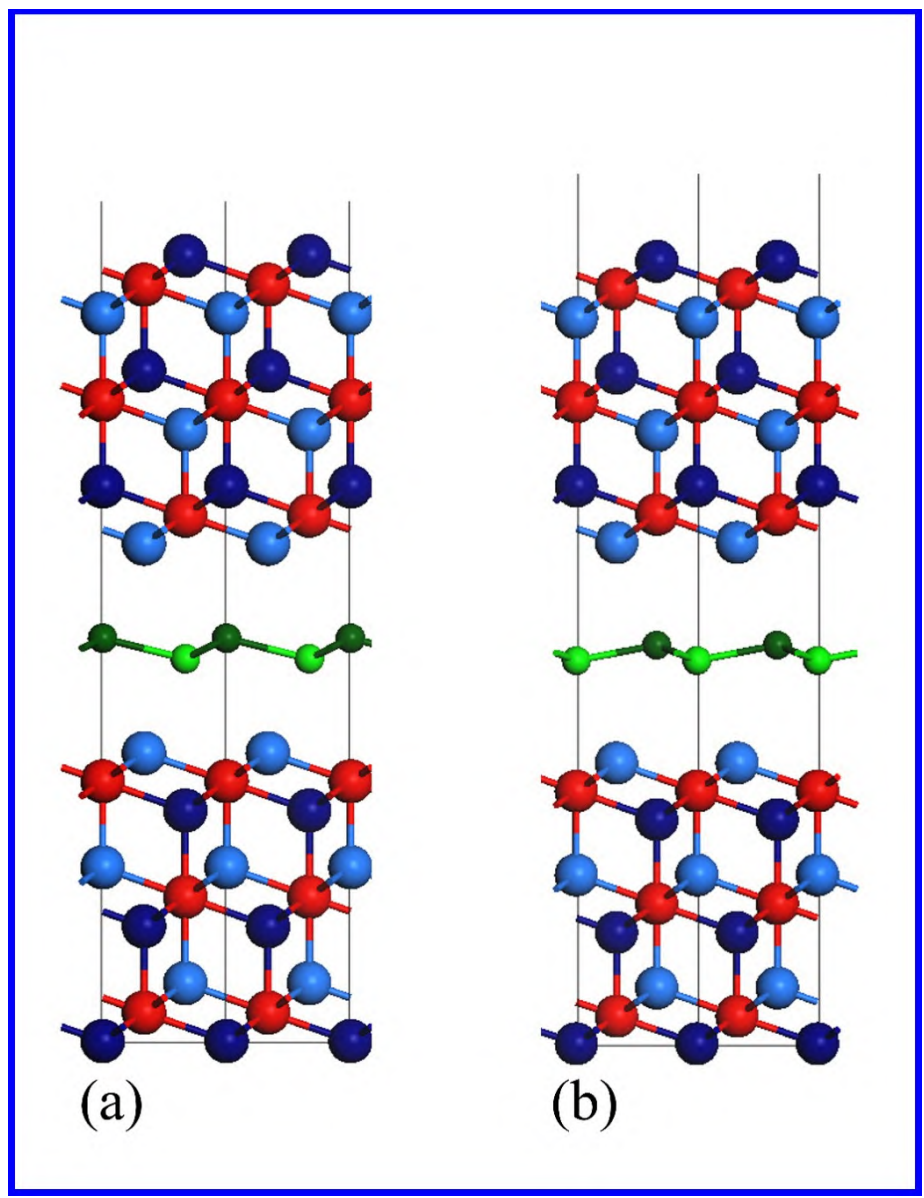

FIG. S8. Geometry for inverse symmetrical sandwich structure CeO$_2$ (111)/silicene/CeO$_2$ (111) corresponding (a) for NA and (b) for NB configuration. Red spheres represent Ce atoms, dark green and light green spheres represent the upper and lower layer Si atoms of silicene, and dark blue and light blue spheres represent O atoms.

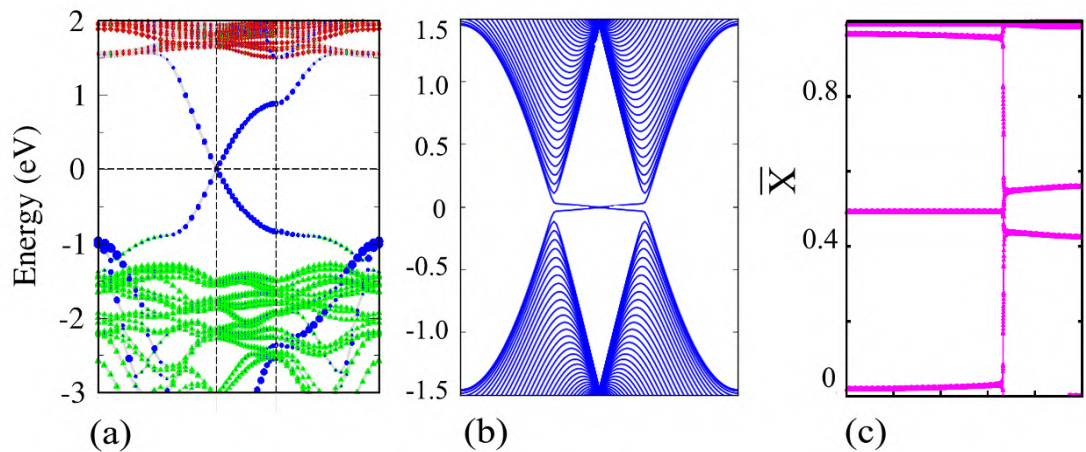

FIG. S9 (a) Band structures of inverse symmetrical sandwich structure $CeO_2$ (111)/Silicene/$CeO_2$ (111) corresponding to NB configuration. Red, green and blue symbols represent the projected band dispersion of Ce, O and Si. (b) Band structure of zigzag nanoribbon of silicene on $CeO_2$ (111) corresponding to NB trilayers. (c) The evolution of the charge centers of the Wannier functions of NB trilayers, implying the $Z_2$ invariant is one.

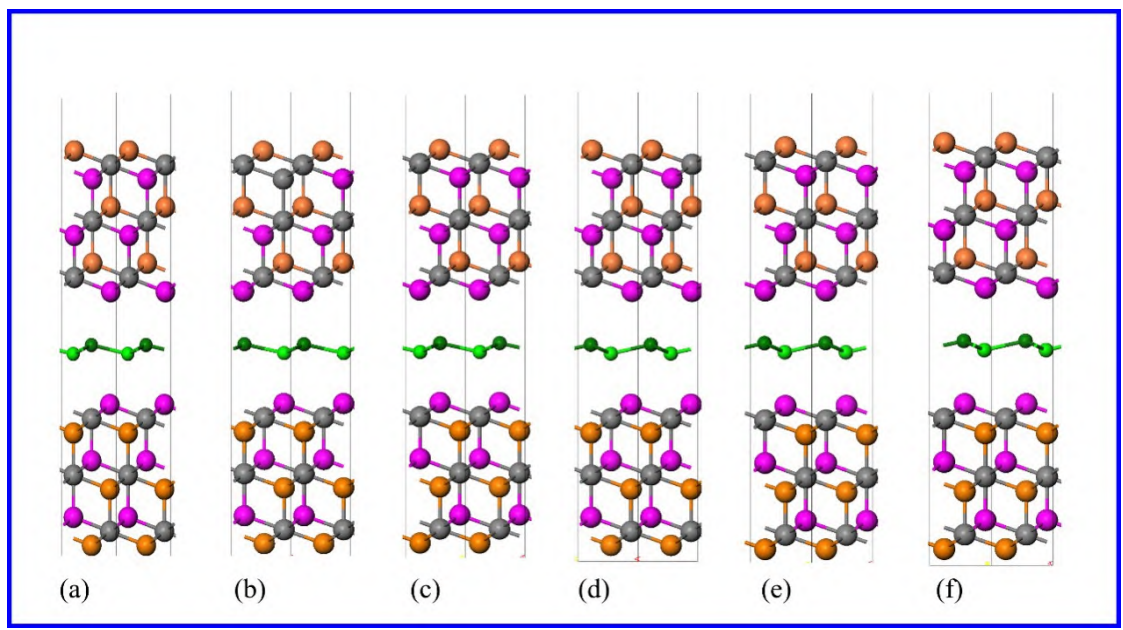

FIG. S10. Geometry for six different configurations of CaF$_2$ (111)/silience/CaF$_2$(111). Gray spheres represent Ca atoms, dark blue and light blue spheres represent the upper and lower layer Si atoms of silicene, and pink and orange spheres represent F atoms.

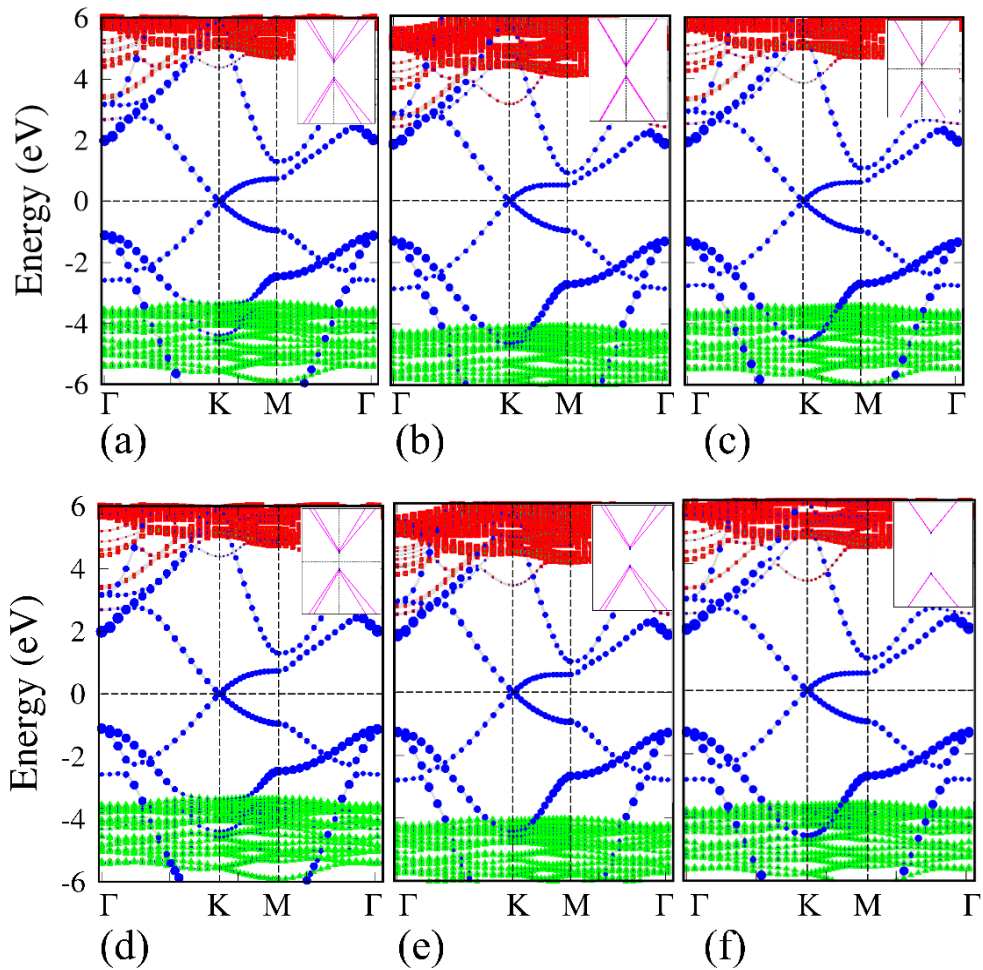

FIG. S11 (a-f) Band structures of silicene/$CaF_2$(111) corresponding to six configurations (a-f) in Fig. S3. Red, green and blue symbols represent the projected band dispersion of Ca, F and Si.

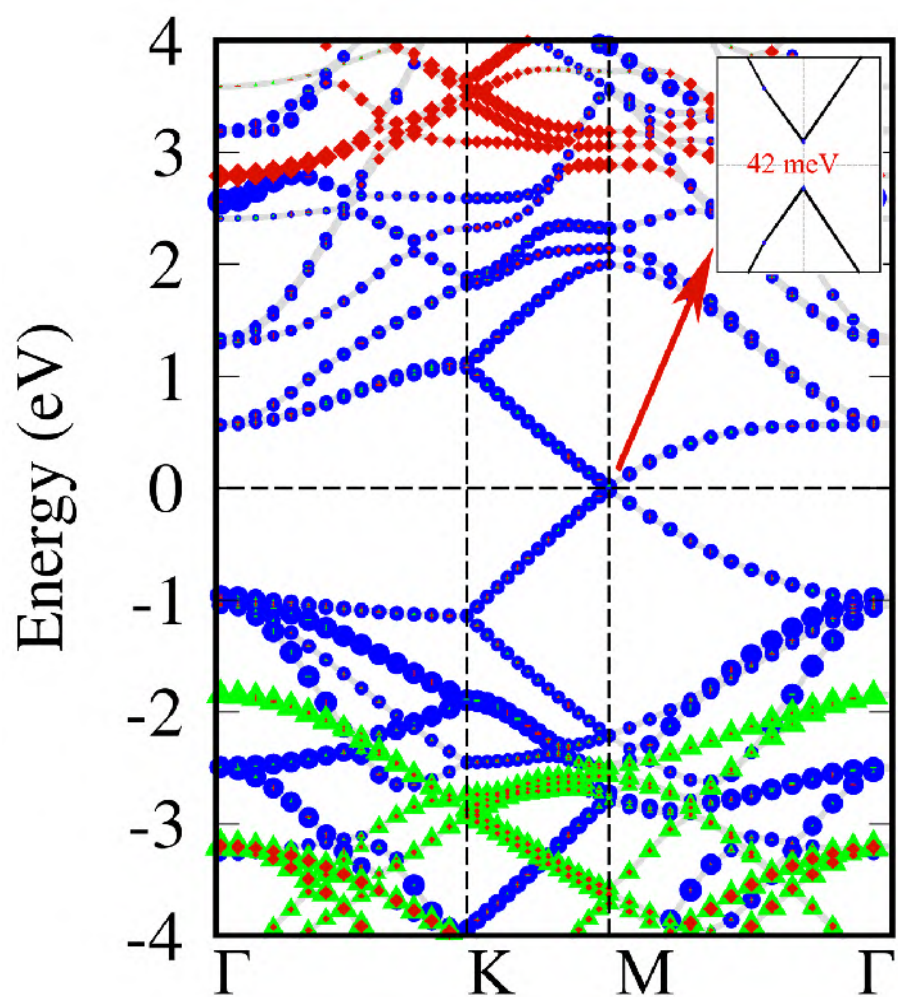

FIG. S12 Band structure for silicene/BN. Red, blue and green symbols represent the projected band dispersion of B, N and Si.

TABLE SI. Silicene on the CaF$_2$ (111) surface. The E$_b$ is the binding energy per Si atom. The E$_g$ is the gap calculated at the K point. The Hamiltonian parameters defined in Eqs. (1) are given in meV. The b is the separation between the upper and lower layer silicene. The d is the interfacial distance between silicene and substrate.

|   | E$_b$(meV) | E$_g$(meV) | $\gamma_m$ | $\gamma_{so}$ | $\gamma_{R1}$ | $\gamma_B$ | b (Å) | $v_F$ ($10^5$m/s) | d (Å) |
|---|---|---|---|---|---|---|---|---|---|
| a | 270.2 | 44.4  | 21.87  | 0.33 | 0.20 | -0.12 | 0.47 | 5.38 | 2.75 |
| b | 249.7 | 253.7 | 125.97 | 0.88 | 1.26 | -0.12 | 0.50 | 5.36 | 2.85 |
| c | 207.9 | 158.1 | 78.17  | 0.88 | 0.49 | -0.07 | 0.43 | 5.48 | 3.08 |
| d | 237.3 | 91.7  | 45.40  | 0.45 | 0.35 |  0.05 | 0.43 | 5.46 | 2.95 |
| e | 272.1 | 379.9 | 189.02 | 0.93 | 2.67 | -0.07 | 0.51 | 5.41 | 2.76 |
| f | 219.3 | 102.2 | 50.12  | 0.98 | 0.59 | -0.12 | 0.45 | 5.47 | 2.99 |